\documentclass{ws-ijmpc}

\begin{document}

\markboth{Nuno Crokidakis and Lucas Sigaud}
{Crime and COVID-19 in Rio de Janeiro: How does organized crime shape the disease evolution?}


\catchline{}{}{}{}{}

\title{Crime and COVID-19 in Rio de Janeiro: How does organized crime shape the disease evolution?}

\author{Nuno Crokidakis and Lucas Sigaud}

\address{Instituto de F\'isica, Universidade Federal Fluminense \\
Niter\'oi, Rio de Janeiro, Brazil\\
nuno@mail.if.uff.br}

\maketitle

\begin{history}
\received{Day Month Year}
\revised{Day Month Year}
\end{history}

\begin{abstract}
The city of Rio de Janeiro is one of the biggest cities in Brazil. Drug gangs and paramilitary groups called \textit{mil\'icias} control some regions of the city where the government is not present, specially in the slums. Due to the characteristics of such two distinct groups, it was observed that the evolution of COVID-19 is different in those two regions, in comparison with the regions controlled by the government. In order to understand qualitatively those observations, we divided the city in three regions controlled by the government, by the drug gangs and by the \textit{mil\'icias}, respectively, and we consider a SIRD-like epidemic model where the three regions are coupled. Considering different levels of exposure, the model is capable to reproduce qualitatively the distinct evolution of the COVID-19 disease in the three regions, suggesting that the organized crime shapes the COVID-19 evolution in the city of Rio de Janeiro. This case study suggests that the model can be used in general for any metropolitan region with groups of people that can be categorized by their level of exposure.


\end{abstract}

\ccode{PACS Nos.: 87.23.Ge, 89.20.-a, 89.65.-s, 89.75.Fb}

\section{Introduction}

The evolution of epidemics is one of the most dangerous problems for a society and its spreading mechanisms have been the source of interest to the exact sciences' scientific community for over a century \cite{kermack}. The humanity faced severe pandemics during its evolution, like the Spanish flu in 1917, the Honk Kong flu (H3N2) of 1968 and the swine flu (H1N1) in 2009. Several efforts were done since the 70's in order to understand the mathematical evolution and spreading of diseases \cite{anderson,bailey}.

The recent outbreak of the COVID-19 pandemic and its effect on society in different countries and its many diverse aspects are now the object of a long list of different studies on the subject of epidemic modelling. An immense and not always reliable amount of data has been produced over the past year regarding infection rates, contagion levels, population-exposure-related contamination, age-related fatality rates, statistical spreading, quarantine efficiency, death tolls and mortality rates by population; these have been exhaustively given worldwide coverage by the press, health organizations and data banks \cite{OMS,worldometer} and we will not delve further here in details regarding these parameters.

For many different diseases, compartmental models focusing on a population divided into Susceptible (\textbf{S}), Infected (\textbf{I}) and Recovered (\textbf{R}) individuals, namely SIR, have been used extensively. It often employs mean-field approximation for the individuals network which, albeit unrealistic, was able to model many different aspects of epidemic behaviour, including the necessity of artificial immunity through vaccination in order to contain the spreading of the disease \cite{andersonmay}. Taking into account the pre-symptomatic, symptomatic and asymptomatic possibilities for COVID-19 infection and spreading, SIR models fail because of two important issues: i) the possibility of pre-symptomatic and asymptomatic contagion through individuals which are not known as patients (i.e. belonging to the \textbf{I} compartment) led to social isolation and quarantine measures throughout the world; and, in consequence, ii) different levels of social isolation among groups coexisting in the same geographical area make the mean-field approximation inadequate for COVID-19 contagion behaviour \cite{odagaki1}.

In order to tackle the first of these two issues, many groups proposed different compartmental modelling of the COVID-19 pandemic spreading using SIQR, with the introduction of a fourth class of individuals, namely the Quarantined ones (\textbf{Q}) \cite{odagaki1,odagaki2,nuno2,elfatini,hackbart,silvio}. The introduction of the \textbf{Q} population is nothing new, however, and quarantine effects on epidemic models have been studied in this manner for a few decades \cite{hethcote}. Nonetheless, SIQR models have been useful in modelling the COVID-19 spreading \cite{nuno2} and, for example, predicting phenomena such as the so-called second wave incidences \cite{odagaki1}. Yah and Li have taken a further step and included age as a novel parameter that influences the connections between the different compartments in a SIQR model \cite{idade}. El Fatini \textit{et al}. \cite{elfatini} have included vaccination strategies employed in previous models \cite{andersonmay} in the very early stages of the COVID-19 onset, while Hackbart \cite{hackbart} has proposed models including mass testing and selected quarantined individuals based on these testing parameters, in order to target infected individuals to reduce contagion rise.

On the other hand, the second issue listed above needs further addressing. Odagaki included the fluctuation of social isolation policies in Japan and how it was obeyed by citizens in order to explain the crescent three-wave pattern that has arisen there over the course of the pandemic \cite{odagaki2}, but still kept the compartmental model with mean-field approximation. Weisbuch attempts to circumvent the limitations of the mean-field approximation by including a mobile initial population at both city and countryside in order to simulate urban exodus and its influence on the COVID-19 spreading \cite{weisbuch}, and then including crossover parameters linking interchange between city and countryside populations. Although it is an oversimplification of the dynamics of evolving networks \cite{zhangetal}, it is an efficient move towards including different types of population in the same model. Kraemer \textit{et al.} have studied the consequences of human mobility and social isolation policies in China during the COVID-19 first months and found strong correlation patterns \cite{kraemer}.

In this work, we aim to address the question of different patterns of social isolation and contact in a population confined to a single geographical location, in order to model groups of individuals either with different degrees of compliance to the same social isolation policy or subjected to different social isolation policies. In order to have numerical data to which to compare the proposed model, we take the particular situation of Rio de Janeiro as a case study, which has been the object of previous works \cite{nuno1}. We chose Rio de Janeiro city because of its very particular subdivision into different areas, as is more thoroughly explained elsewhere \cite{arias,magaloni,crime_e_covid}. In short, although in principle the entire city is subjected to the regular official municipal, state and federal policies, two other groups, with divergent governance systems, retain political and enforcement control over many different sub-regions of Rio de Janeiro's metropolitan area, namely drug trafficking gangs and the paramilitary groups called \textit{mil\'icias}. While drug trafficking gangs exert stay-at-home regulations in their controlled areas, since their financial activities rely on underground commerce and minimal governmental interference, \text{mil\'icias}, operated by police and ex-police officers as a parallel power, force local populations to not stay at home, in order to maintain their commercial tax and protection monetary charges.  This behavior in different slums has been monitored using, for example, credit card and mobile cellphones usage data \cite{slums_covid} and an illustrative map of the different regions is shown by Bruce \textit{et al.} \cite{crime_e_covid}, but it is not reproduced here because we do not take the geographical subdivisions into account in our model. Therefore, in the same metropolitan area there is an interchange of individuals subdivided into three different but connected populations, each with its own exposure parameter, which is related to the level of social distancing that the individuals for each group adopt. In other words, the exposure coefficients are related to the degree of exposure that each individual both is subjected to other individuals and subject other individuals to him/herself. We aim to verify if the distinct contagion rates observed by the authors \cite{crime_e_covid} are a direct consequence of different exposure patterns.

In order to try to accomplish this, we refrain from using the SIQR model, since, in general, our goal is to address the effects of exposure patterns on the spread of COVID-19, and, in particular for our case study, quarantined individuals data are very hard to obtain in both drug trafficking gangs and \textit{mil\'icias} controlled regions. We opt instead for the SIRD model, which includes a compartment for Dead (\textbf{D}) individuals and is very useful when death rates data are more readily available, which is the case for COVID-19 both in general \cite{sird_Brazil,sird_india} and in our particular case study, where death rates due to COVID-19 are reported to be higher for \textit{mil\'icia}-controlled areas and lower for drug-trafficking-gang-controlled areas, with respect to the rates observed for regions under the direct influence of neither type \cite{crime_e_covid}.


\section{Model}

Let us recall the Susceptible-Infected-Recovered-Dead (SIRD) model \cite{kermack}. The population is divided in four compartments, and the equations for the evolution of such compartments are
\begin{eqnarray} \label{eq1}
\frac{dS}{dt} & = & -\beta\,S\,I  \\ \label{eq2}
\frac{dI}{dt} & = & \beta\,S\,I - \gamma\,I - \delta\,I \\ \label{eq3}
\frac{dR}{dt} & = & \gamma\,I \\ \label{eq4}
\frac{dD}{dt} & = & \delta\,I
\end{eqnarray}
\noindent
where $S, I, R$ and $D$ denote the fractions of Susceptible, Infected, Recovered and Dead individuals, respectively. The parameters $\beta$, $\gamma$ and $\delta$ denote the infection, recovery and death probabilities, respectively.

As discussed in the Introduction, we are interested to verify if the hypothesis pointed by the authors in Ref. \cite{crime_e_covid} is plausible. In other words, if the restrictions to stay at home imposed by the drug trafficking gangs in some regions of the city and the pressure to not stay at home in areas dominated by the \textit{mil\'icias} influenced the COVID-19 spreading in distinct areas of Rio de Janeiro. For this purpose, we divided the Rio de Janeiro city in three regions, namely:
\begin{itemize}
\item \textbf{1}: areas controlled by the government
\item \textbf{2}: areas controlled by drug trafficking gangs
\item \textbf{3}: areas controlled by \textit{mil\'icias} 
\end{itemize}  

In each of such regions, we consider a Susceptible-Infected-Recovered-Dead (SIRD) model. In addition, we consider in the models' equations the exposure restrictions (in the case of drug gangs' areas) or the absence of such restrictions (in the case of  \textit{mil\'icias}). In such case, we consider three additional parameters to the SIRD model: the coefficients related to the population's exposure in each region, namely $C_1, C_2$ and $C_3$ for regions $1, 2$ and $3$, respectively. Those coefficients take into account the couplings of the population of a given region with the population of the other regions.

Thus, the model is defined through the equations
\begin{eqnarray} \label{eq5}
\frac{dS_u}{dt} & = & -\beta\,S_u\,C_u\sum_{v=1}^{3}C_v\,I_v  \\ \label{eq6}
\frac{dI_u}{dt} & = &  \beta\,S_u\,C_u\sum_{v=1}^{3}C_v\,I_v - \gamma\,I_u - \delta\,I_u \\ \label{eq7}
\frac{dR_u}{dt} & = & \gamma\,I_u \\ \label{eq8}
\frac{dD_u}{dt} & = & \delta\,I_u 
\end{eqnarray}
\noindent
where $S_u, I_u, R_u$ and $D_u$ denote the fractions of Susceptible, Infected, Recovered and Dead individuals, respectively, in each region $u$, with $u=\{1,2,3\}$. The terms $S_u\,C_u\,C_v\,I_v$ in Eqs. \eqref{eq5} and \eqref{eq6} introduce the above-mentioned coupling among the regions. In other words, the coupling terms correspond to the infection of individuals in a given location $u$ by infected individuals \cite{weisbuch,pires} of all the three locations ($v=\{1,2,3\}$) that move along the city. Notice that we do not restrict the $C_u$ values to the range $0<C_u<1$ since, strictly speaking, they are not probabilities -- they are related to the degree of exposure of individuals and they depend on the specific area of the city such individuals live. For simplicity, we considered homogeneous probabilities $\beta$, $\gamma$ and $\delta$ for the three regions.


\section{Results}

We considered the initial conditions $R_u(0)=D_u(0)=0$ and $I_u(0)=0.001$ for $u=\{1,2,3\}$. The normalization condition is obeyed at each time step, i.e., $S(t)=1-I(t)-R(t)-D(t)$, where $S(t)=S_1(t)+S_2(t)+S_3(t), I(t)=I_1(t)+I_2(t)+I_3(t), R(t)=R_1(t)+R_2(t)+R_3(t)$ and $D(t)=D_1(t)+D_2(t)+D_3(t)$. In this case, the initial condition for the susceptible populations is $S_1(0)=S_2(0)=S_3(0)=S(0)/3$.

Since we have distinct behaviors regarding the social isolation in the three regions, we have $C_2 < C_1 < C_3$, i.e., there is less exposure in drug trafficking regions (area 2) and more in \textit{mil\'icia} controlled ones (area 3). As we are interested in the qualitative description of the results obtained in \cite{crime_e_covid}, and the parameters $C_1, C_2$ and $C_3$ are the novelty of the model (as well as the division in three regions), we fixed the transition probabilities $\beta=0.32$, $\gamma=0.045$ and $\delta=0.054$, based on estimates for COVID-19 parameters \cite{nuno1,lancet_days}.

\begin{figure}[t]
\begin{center}
\vspace{3mm}
\includegraphics[width=0.45\textwidth,angle=0]{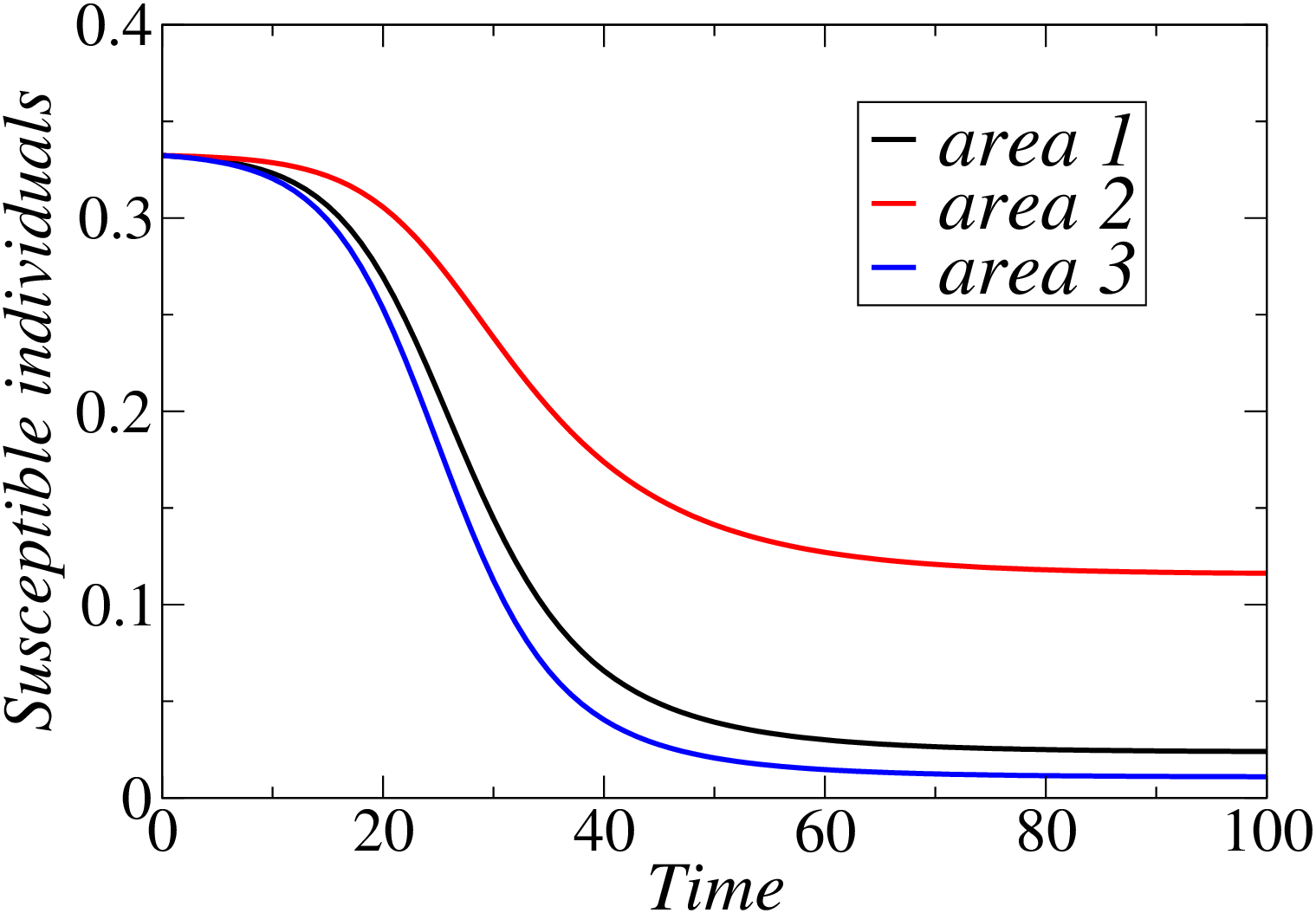}
\hspace{0.2cm}
\includegraphics[width=0.46\textwidth,angle=0]{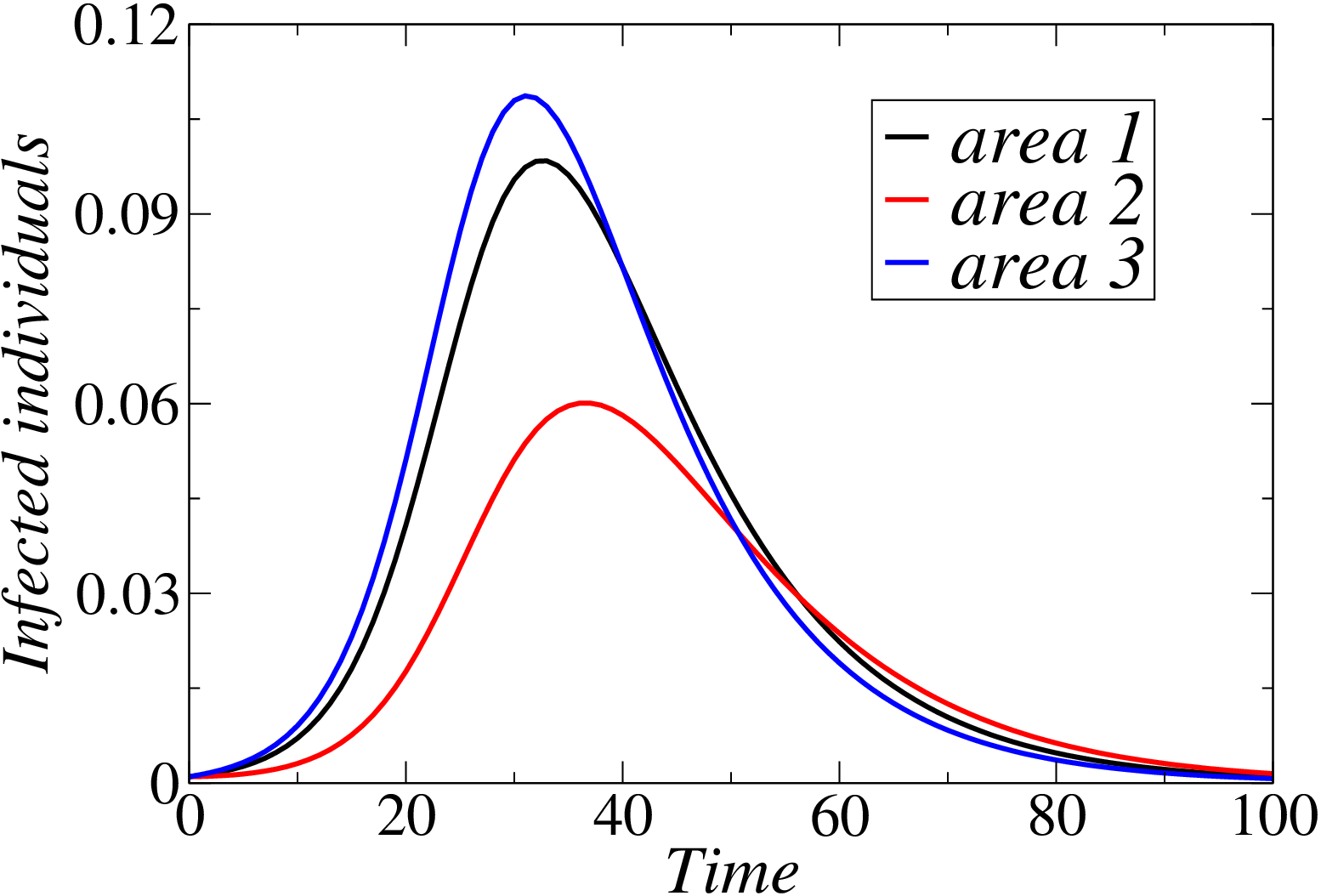}
\\
\vspace{1.0cm}
\includegraphics[width=0.45\textwidth,angle=0]{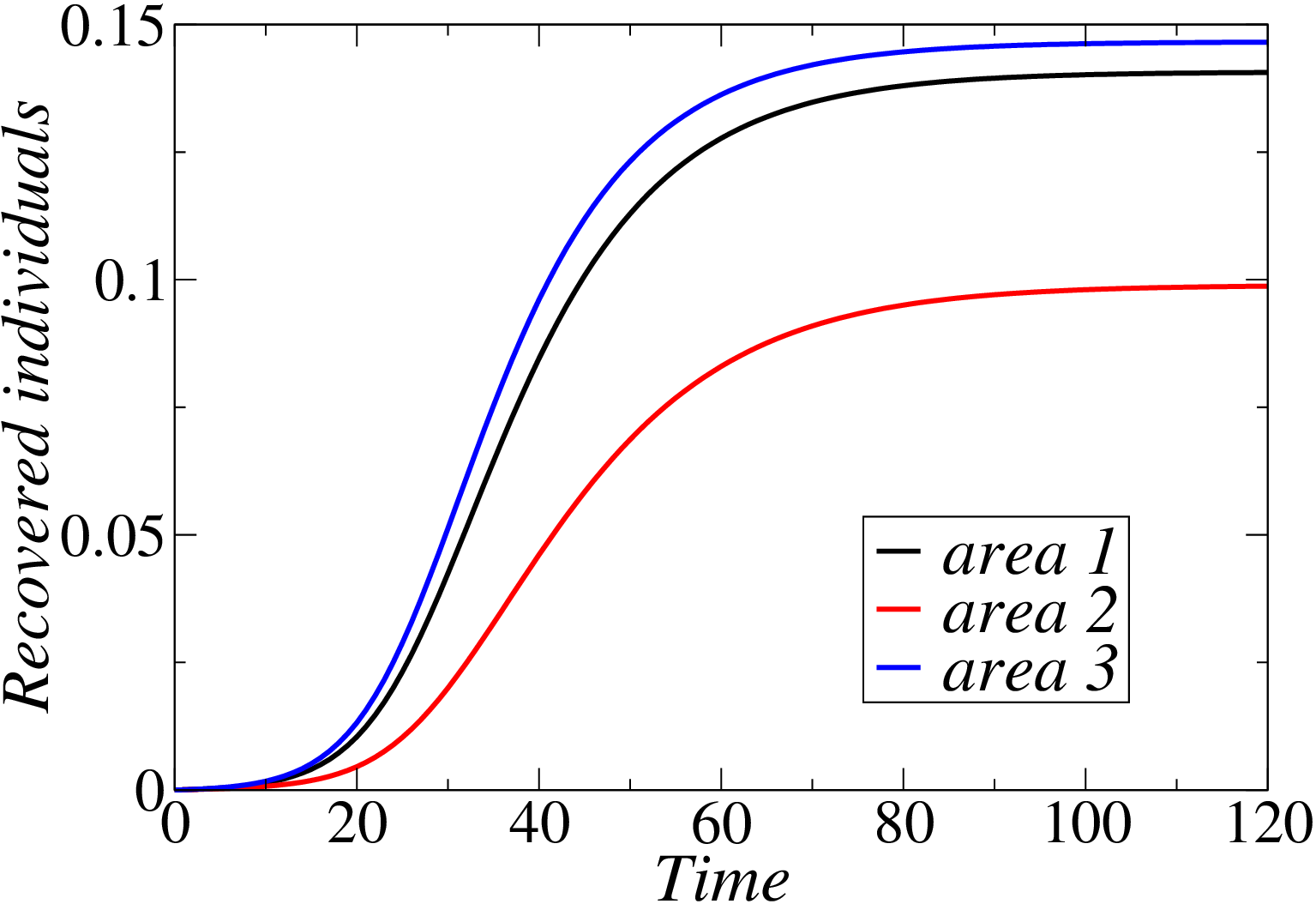}
\hspace{0.2cm}
\includegraphics[width=0.45\textwidth,angle=0]{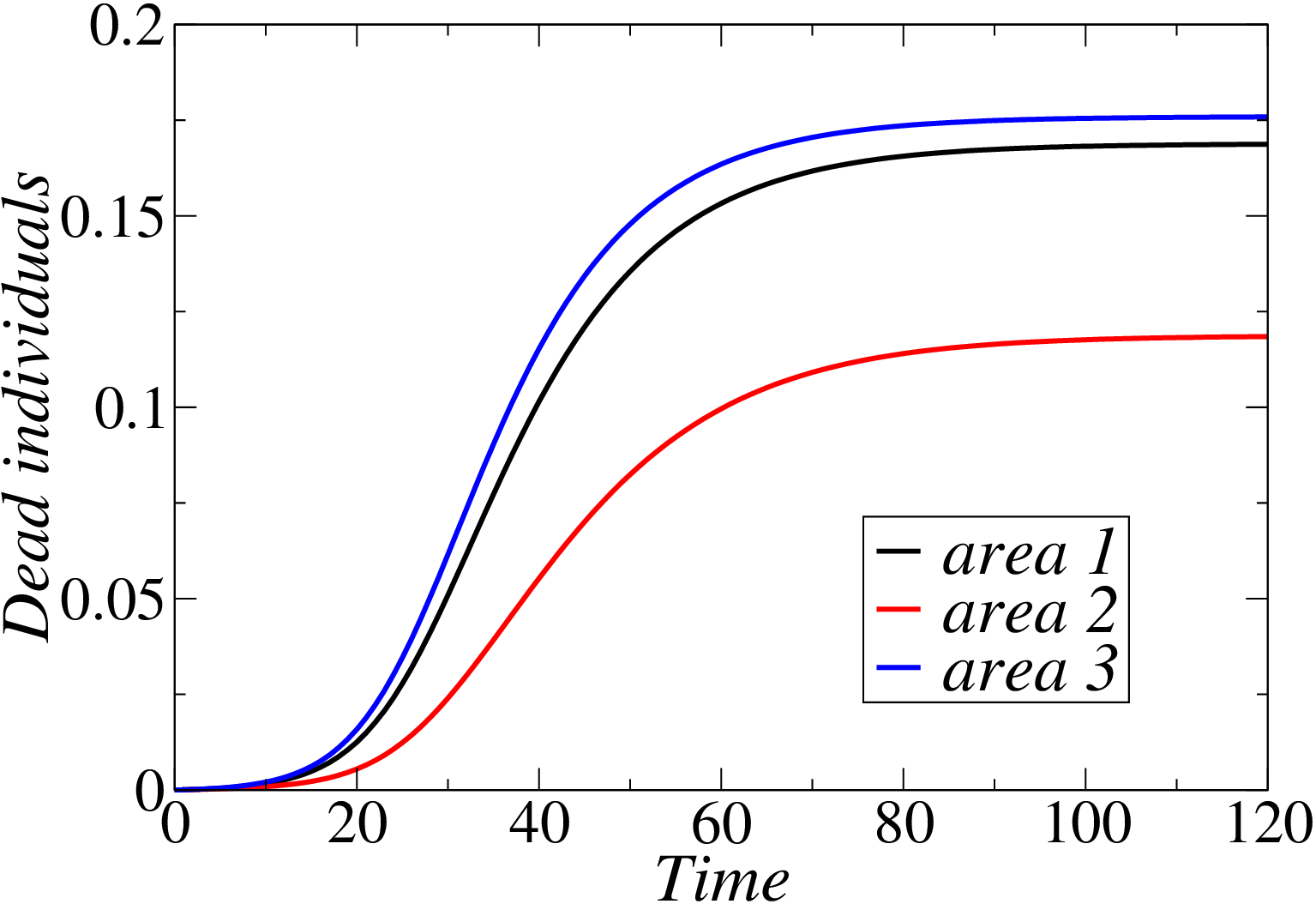}
\end{center}
\caption{Fractions of Susceptible, Infected, Recovered and Dead individuals for the areas 1 (control - black curve), 2 (drug gangs - red curve) and 3 (mil\'icias - blue curve) as functions of time for $C_1=1.0$, $C_2=0.4$ and $C_3=1.3$. We can see distinct time evolutions for the 3 regions due to distinct individuals' degrees of exposure.}
\label{fig1}
\end{figure}

In Fig. \ref{fig1} we exhibit the fractions of Susceptible, Infected, Recovered and Dead individuals for the three areas 1, 2 and 3 as functions of time for $C_1=1.0$, $C_2=0.4$ and $C_3=1.3$, obtained from the numerical integration of Eqs. \eqref{eq5} - \eqref{eq8}. Since the exposure coefficients are arbitrary parameters, we chose as unity the one for government controlled areas and modified the other two accordingly. We can see that if we consider the coupling among the 3 regions, together with the distinct exposure coefficients $C_u$ ($u=\{1,2,3\}$), the model leads to distinct evolutions in each region $u$. In particular, one can see that the outbreak for the infections is more pronounced in the region 3 dominated by the mil\'icias. On the other hand, one observe the flattening of the curve of infections in region 2, dominated by the drug gangs. Such difference among the three regions leads to distinct evolutions of the number of the other three compartments $S, R$ and $D$. Specifically, the number of deaths is higher in region 3 and lower in region 2, in comparison with the region 1 controlled by the State. These results suggest that the scenario observed in \cite{crime_e_covid}, i.e. distinct evolutions of the infections and deaths in the three regions of Rio de Janeiro, is indeed due to the distinct exposure patterns imposed by drug trafficking gangs and mil\'icias, in such two regions (2 and 3) that are not controlled by the State.

\begin{figure}[t]
\begin{center}
\vspace{3mm}
\includegraphics[width=0.45\textwidth,angle=0]{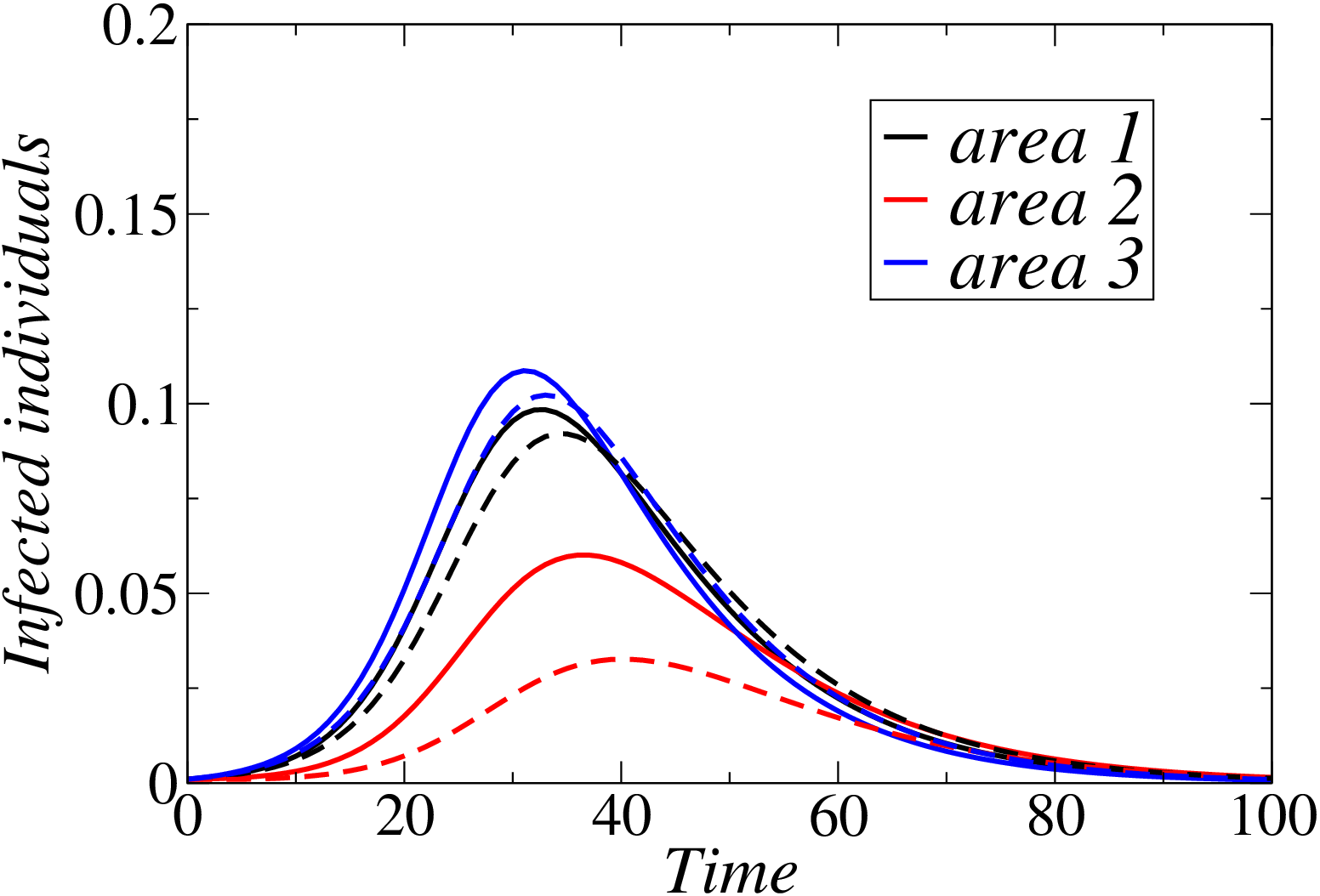}
\hspace{0.2cm}
\includegraphics[width=0.45\textwidth,angle=0]{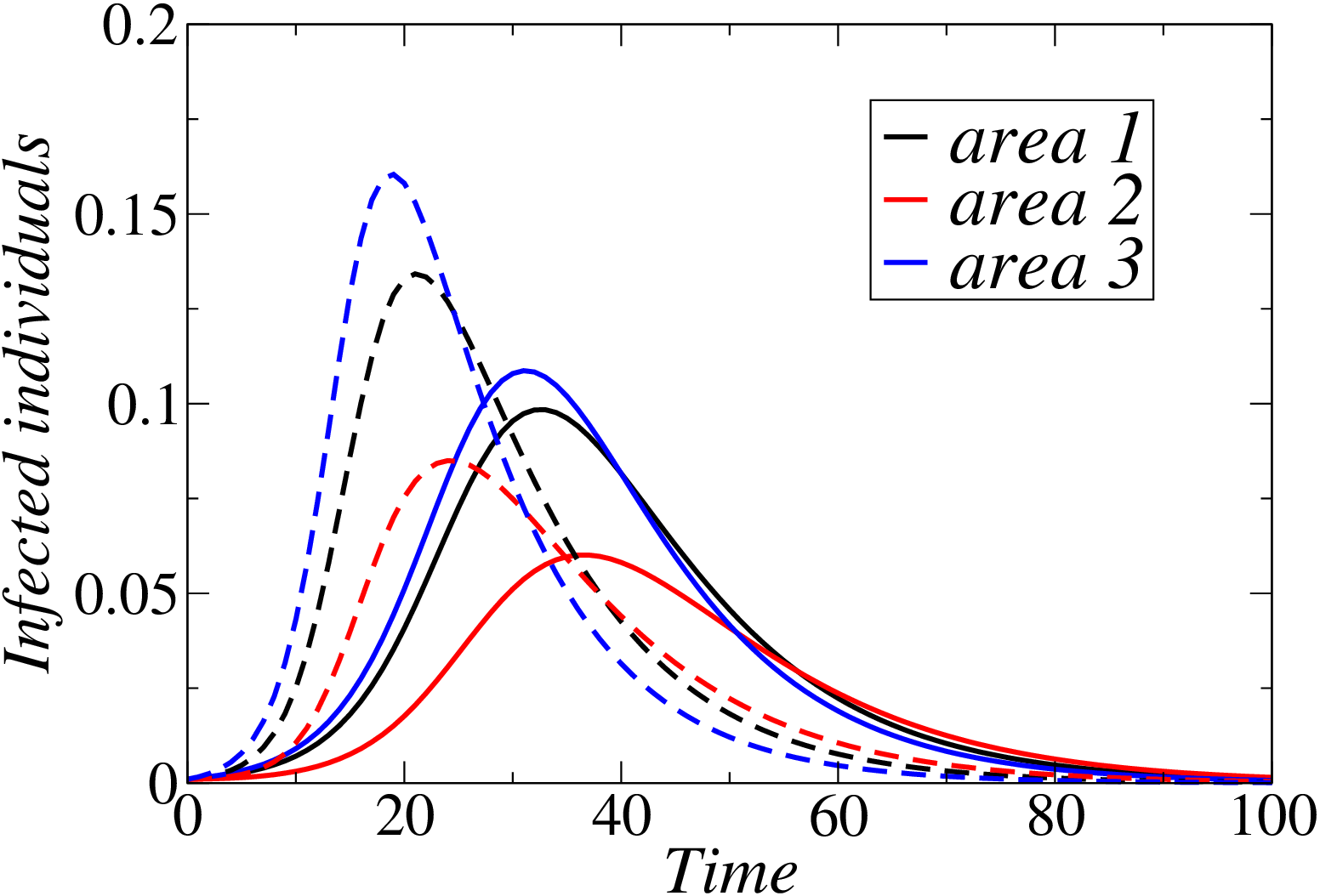}
\end{center}
\caption{Comparative fractions of Infected individuals for the areas 1 (control), 2 (drug gangs) and 3 (mil\'icias) as functions of time for fixed $C_1=1.0$. Left panel: fixed $C_3=1.3$ with $C_2=0.4$ (full lines) and $C_2=0.2$ (dashed lines). Right panel: fixed $C_2=0.4$ with $C_3=1.3$ (full lines) and $C_3=1.8$ (dashed lines).}
\label{fig2}
\end{figure}

In Fig. \ref{fig2} we exhibit comparative results for the fraction of Infections in the three regions. We fixed the standard exposure coefficient for region 1 as $C_1=1.0$, and present two distinct comparisons. First, we consider the curves for fixed $C_3=1.3$ (as in Fig. \ref{fig1}) and distinct values of $C_2$ (Fig. \ref{fig2}, left-hand side), namely $C_2=0.4$ (full lines) and $C_2=0.2$ (dashed lines). One can see that lowering the exposure coefficient for the more social isolated group (the one controlled by drug trafficking gangs in our case study) considerably lowers the infection rates for individuals belonging to this group, while maintaining the other two groups fairly unchanged. On the other hand, in the right-hand side panel we consider the curves for fixed $C_2=0.4$ (as in Fig. \ref{fig1}) and distinct values of $C_3$, namely $C_3=1.3$ (full lines) and $C_3=1.8$ (dashed lines).
It is very interesting to see that in this case, when we change the degree of exposure of the less social isolated group (the \textit{mil\'icia} controlled one, in our case study), the infected population of all three groups increase considerably, and roughly by the same proportion. This clearly indicates that, in a city divided in three different groups, with different degrees of exposure, increasing the social distancing among individuals that already have a lower degree of exposure, albeit lowering the infection rates of this particular group, does not change the overall scenario: the two other groups, which were already responsible for the larger share of infected individuals are only slightly affected by it. Meanwhile, if we increase the degree of exposure (i.e. decrease the social distancing) of individuals which were not respecting social isolation policies in the first place, the consequences for the entire population are indeed drastic. It is worthwhile to mention that this does not refer to our case study alone - in any populated area where one is able to subdivide the inhabitants into groups with three different degrees of exposure, this conclusion holds. Moreover, since we assumed a mean-field-like approximation modulated by the exposure coefficients, we did not take into account the geographical properties of Rio de Janeiro City and its controlled regions; therefore, this aspect takes a more general form.

From Fig. \ref{fig2} one can also see some of the more general infection behaviors that always appear when looking at epidemic models: the flattening (height decrease and enlargement) of the infection curves over time for higher social isolation patterns and a shorter pandemic time-frame with increased infection (and, consequently, death) rates for lower social distancing. It is meaningful that these behaviors are reproduced as well in a population subdivided as we presented here, with these differences showing simultaneously between the three different groups; i.e. group 2 has flatter infection rate curve over time than groups 1 and 3. From the changes shown in Fig. \ref{fig2} for the three curves when different values for parameters $C_2$ or $C_3$ were chosen, one can infer that the existence of a group which does not respect social distancing measures has a very strong effect on individuals that are not under a strict social distancing policy.


\section{Final Remarks}   

In this work we considered a  Susceptible-Infected-Recovered-Dead (SIRD) model in order to study the evolution of COVID-19 in the Rio de Janeiro city. Taking into account that the organized crime dominates several neighborhoods of the city, we divided the city in three areas, namely the areas controlled by the government (area 1), by the drug trafficking gangs (area 2) and by the paramilitary groups called \textit{mil\'icias} (area 3). Taking into account that the behavior of individuals in each region is distinct, determined by the government in area 1 and by the organized crime in areas 2 and 3, we considered distinct exposure coefficients for the three regions. In addition, we coupled the 3 regions through these exposure coefficients, considering a not-independent SIRD model in each region.

At this point, some limitations can be discussed. First of all, we considered a mean-field-like approach, where each individual can interact with all others. In this case, spatial features were not considered in the model. Our work does not consider explicitly an upper bound for the capacity of the healthcare system. Under-reporting is another feature that is not modeled here. We have also assumed permanent immunity, i.e., after recovering from COVID-19 individuals cannot be infected anymore. In addition, as we are interested in explaining a phenomenon observed in the beginning of the epidemic evolution, we did not implement other containment measures besides the exposure restrictions imposed by each controlling agent. 

It is important to say that our target was not to describe exactly the evolution of COVID-19 in Rio de Janeiro city, but we are interested in the qualitative description of an observed phenomenon, namely the distinct evolution of the number of cases and deaths in the three distinct regions, as observed by economists in recent works \cite{crime_e_covid,slums_covid}. In such a case, despite the above-mentioned limitations, the model presented here is capable to explain in a simple way how the organized crime shaped the COVID-19 evolution in Rio de Janeiro.

Furthermore, since we employ a mean-field-like approach, the geographical features and subdivisions of our case study were not taken into account, which in a certain way can be actually considered a feature instead of a liability: it shows, in a first-order approximation, the consequences of the epidemic progression in individuals belonging to different exposure patterns. It can be inferred from our results that a group of individuals that respects social distancing policies has less impact on the overall pandemic picture -- besides, of course, lowering their own infection and death risks -- than a group of individuals that presents higher degrees of exposure. The increase of exposure in the latter group considerably enhances the infection rate, while the decrease of exposure in the former, social isolated, group has a much lesser impact on the progression of the disease.

\section*{Acknowledgments}

The authors acknowledge financial support from the Brazilian scientific funding agencies CNPq (Grants 303025/2017-4 and 311019/2017-0) and FAPERJ (Grant 203.217/2017).


\begin{thebibliography}{00}




\bibitem{kermack}
W. O. Kermack, A. G. McKendrick, \textit{A contribution to the mathematical theory of epidemics}, Proceedings of the Royal Society of London, Series A, 115 (772):700–721 (1927).


  
\bibitem{anderson}
R. M. Anderson, R. M. May, \textit{Infectious Diseases of Humans: Dynamics and  Control} (Oxford University Press, Oxford, 1991).

\bibitem{bailey}
N. T. J. Bailey, \textit{The Mathematical Theory of Infectious Diseases and its Application} (Hafner Press, New York, 1975).

\bibitem{OMS}
The WHO official COVID-19 website (https://www.who.int/emergencies/diseases/novel-coronavirus-2019).

\bibitem{worldometer}
The worldometer data bank for COVID-19 statistics (https://www.worldometers.info/coronavirus/)

\bibitem{andersonmay}
R.M. Anderson and R.M. May, \textit{Directly Transmitted Infectious Diseases: Control by Vaccination}, Science 215, 1053 (1982).

\bibitem{odagaki1}
T. Odagaki, \textit{Exact properties of SIQR model for COVID-19}, Physica A 564, 125564 (2021).

\bibitem{odagaki2}
T. Odagaki, \textit{Self‑organized wavy infection curve of COVID‑19}, Scientific Reports 11, 1936 (2021).

\bibitem{nuno2}
N. Crokidakis, \textit{Modeling the early evolution of the COVID-19 in Brazil: Results from a Susceptible–Infectious–Quarantined–Recovered (SIQR) model}, International Journal of Modern Physics C 31, 2050135 (2020).

\bibitem{elfatini}
M. El Fatini, R. Pettersson, I. Sekkak and R. Taki, \textit{A stochastic analysis for a triple delayed SIQR epidemic model with vaccination and elimination strategies}, Journal of Applied Mathematics and Computing 64, 781 (2020).

\bibitem{hackbart}
G. Hackbart, \textit{Heterogeneous SIQR Models with Mass Testing and Targeted Quarantine and the Spread of Infectious Diseases}, Working paper (July, 2020), Available at SSRN: $https://ssrn.com/abstract=3643816$ or $http://dx.doi.org/10.2139/ssrn.3643816$.

\bibitem{silvio}
N. Crokidakis, S. M. D. Queirós, \textit{Probing into the effectiveness of self-isolation policies in epidemic control}, Journal of Statistical Mechanics P06003 (2012).
  

\bibitem{hethcote}
H. Hethcote, M. Zhien and L. Shengbing, \textit{Effects of quarantine in six endemic models
for infectious diseases}, Mathematical Biosciences 180, 141 (2002).

\bibitem{idade}
H. Yah and J. Li, \textit{SIQR dynamics in a random network with heterogeneous connections with infection age}, Journal of Nonlinear Sciences \& Applications 14, 196 (2021).

\bibitem{weisbuch}
G. Weisbuch, \textit{Urban exodus and the dynamics of COVID-19 pandemics}, Physica A 569, 125780 (2021).

\bibitem{zhangetal}
X. Zhang, C.Z. Shan Jin and H. Zhu, \textit{Complex dynamics of epidemic models on adaptive networks}, Journal of Differential Equations 266, 803 (2019).

\bibitem{kraemer}
M. U. G. Kraemer \textit{et. al.}, \textit{The effect of human mobility and control measures on the COVID-19 epidemic in China}, Science 10.1126/science.abb4218 (2020).

\bibitem{nuno1}
N. Crokidakis, \textit{COVID-19 spreading in Rio de Janeiro, Brazil: Do the policies of social isolation really work?}, Chaos, Solitons \& Fractals 136, 109930 (2020).

\bibitem{arias}
E. D. Arias and C. D. Rodrigues, \textit{The myth of personal security: Criminal gangs, dispute resolution, and identity in Rio de Janeiro’s favelas}, Latin American Politics And Society 48, 53 (2006).

\bibitem{magaloni}
B. Magaloni, E. Franco-Vivanco and V. Melo, \textit{Killing in the slums: Social
order, criminal governance, and police violence in Rio de Janeiro}, American Political Science Review 114, 552 (2020).

\bibitem{crime_e_covid}
R. Bruce,  A. Cavgias, L. Meloni, \textit{Filling the Void? Organized Crime and COVID-19 in Rio De Janeiro}, Working paper (August 21, 2020), Available at SSRN: $https://ssrn.com/abstract=3678840$ or $http://dx.doi.org/10.2139/ssrn.3678840$.

\bibitem{slums_covid}
L. Brotherhood, T. Cavalcanti, D. Da Mata, C. Santos, \textit{Slums and Pandemics} (August 2, 2020), Available at SSRN: $https://ssrn.com/abstract=3665695$ or $http://dx.doi.org/10.2139/ssrn.3665695$   

\bibitem{sird_Brazil}
R. M. da Silva, \textit{Using the SIRD model to characterize the COVID-19 spreading in the states of Paraná, Rio Grande do Sul, and Santa Catarina}, SciELO Preprints, $DOI: https://doi.org/10.1590/SciELOPreprints.764$

\bibitem{sird_india}
S. Chatterjee, A. Sarkar, S. Chatterjee, M. Karmakar, R. Paul, \textit{Studying the progress of COVID-19 outbreak in India using SIRD model}, Indianian Journal of Physics (2020), $https://doi.org/10.1007/s12648-020-01766-8$


\bibitem{pires}
M. A. Pires  \textit{et. al.}, \textit{What is the potential for a second peak in the evolution of SARS-CoV-2 in emerging and developing economies? Insights from a SIRASD model considering the informal economy}, arXiv:2005.09019 (2020).

  
\bibitem{lancet_days}  
Fei Zu \textit{et. al.}, \textit{Clinical course and risk factors for mortality of adult inpatients with COVID-19 in Wuhan, China: a retrospective cohort study}, Lancet 395, Issue 10229, 1054-1062 (2020).


  

  
  

\end{thebibliography}
\end{document}